\documentclass[12pt]{article}
\usepackage{lingmacros}
\usepackage{tree-dvips}
\usepackage{graphicx} 
\usepackage{setspace}
\usepackage{makeidx}
\usepackage[utf8]{inputenc}
\usepackage{fancyhdr}
\usepackage{lipsum}
\usepackage{mathtools,amssymb}
\usepackage{breqn}
\usepackage{float}
\begin{document}
	\title{NMR of Confined Fluids: a Numerical Study using COMSOL with application in Petrophysics}
	\maketitle
	\begin{center}
	\author{Ivan S. Oliveira \\ Brazilian Center for Research in Physics  \\ Rua Dr. Xavier Sigaud, 150, Rio de janeiro, 22290-180, Brazil \\
	Contact: ivan@cbpf.br}
	\end{center}
	\begin{abstract}
	Nuclear Magnetic Resonance (NMR) is one of the main experimental tools to evaluate the production potential of porous rocks in oil wells. From the relative areas and mean values obtained from relaxation time distribution curves, information about fluid content, porosity and permeability can be obtained.   In this report, a numerical study of pulsed NMR of confined fluids using the software COMSOL (license 9200972) is presented. This is done by solving the Bloch-Torrey equations considering surface relaxivity using the ``flux/source'' boundary conditions for different geometries, in the fast and slow diffusion regimes.  The study is made for a single fluid and for a mixture of two coupled fluids (McConnell equations), first in isolated pores, and then in a ``cylindrical porous plug" containing one single 20$\mu$m spherical pore surrounded by $800 \times 2\mu$m, unconnected, also spherical ones. NMR spectra and transverse relaxation are calculated as a function of pore sizes and coupling strength between fluids, for single $\pi/2$ pulses, followed by the FFT of the NMR signal (FID), and the echo amplitude after a spin-echo pulse sequence, respectively. The simulations show that,  in the fast diffusion regime, pore sizes can be, in principle, distinguished by transverse relaxation, but even in noiseless condition and full control of geometry, only a rough estimate of sizes can be obtained. A number of possible follow-up studies with COMSOL are discussed at the concluding remarks. 
\end{abstract}
\index{entry}
\printindex
\doublespacing

\section*{Introduction} The best way to begin this report is, perhaps, stating what it is NOT about: it is not a treatise on NMR Petrophysics, a subject far complex and deep. But it is not  a manual of use of COMSOL, either. So, what is it, then? Hopefully it is a motivation for those students or professionals working in the area of Petrophysics (or any subject involving a porous system) and/or NMR, seeking new tools which can be used to help understanding the very rich NMR phenomena inside porous media. The approach here is to exploit the simplest of the NMR experiments to make numerical simulations with COMSOL, also in the simplest of the geometries, as a ``proof of principle'', without  going into the inexhaustible list of existing techniques in NMR. Some possible developments are discussed in the Concluding Remarks.

Nuclear Magnetic Resonance (NMR) is one of the main tools in the investigation of petrophysical properties of porous rocks for the determination of porosity, permeability and fluid content  of natural rocks \cite{ref1}. NMR data, either collected in field (NMR logging), or in laboratory, along with other petrophysical techniques, assist  geophysicists, engineers and technicians in the many million dollars decision of drilling or not an oil well. NMR was decisive, for instance, to the discovery of the Brazilian pre salt oil field. 

The two main References which paved the way of NMR to become an important petrophysical technique are the 1956 paper by H.C. Torrey, which introduced diffusion into the Bloch Equations \cite{ref2}, and  the 1979 work by K.R. Brownstein and C.E. Tarr, which applied the so-called Bloch-Torrey equations to the decay of nuclear magnetization decay (usually characterized by the time constant $T_2$) of confined fluids \cite{ref3}. 

In the rotating reference frame, for a radiofrequency (RF) pulse applied on the $X-Y$ plane with arbitrary phase $\phi$, the so-called Bloch-Torrey equations can be written, for a single fluid component, in the following matrix form  \cite{ref2a}: 

\[
	\frac{\partial}{\partial t}
	\left(\begin{array}{c}
		m_x \\
		m_y \\
		m_z
	\end{array}\right)+
\left(\begin{array}{ccc}
	1/T_2 & -\Delta\omega & \omega_1\sin\phi \\
	\Delta\omega & 1/T_2 & -\omega_1\cos\phi \\
	-\omega_1\sin\phi & \omega_1\cos\phi & 1/T_1
\end{array}\right)
\left(\begin{array}{c}
	m_x \\
	m_y \\
	m_z
\end{array}\right) - \]
\begin{equation}\label{eq1}
-D\nabla^2\left(\begin{array}{c}
	m_x \\
	m_y \\
	m_z
\end{array}\right)=
\left(\begin{array}{c} 
	0 \\
	0 \\
	m_0/T_1
\end{array}\right)
\end{equation} where $m_0$ is the equilibrium magnetization density, $m_k(\vec{r},t)$, $k=x,y,z$, is the $k^{th}$ component of the magnetization density at position $\vec{r}$ and instant $t$, $T_1$ and $T_2$ are, respectively,  spin-lattice and spin-spin relaxation times. $\Delta\omega$ is the frequency offset and $\omega_1$ the coupling strength to the RF field. In the following simulations, unless stated otherwise,  $\phi=0$ will be set, which corresponds to RF pulses applied along the $X$ direction in the rotating frame. However, the general form of Eq.(\ref{eq1}) can be used, for instance, to study the effects of phase cycling on the NMR signal.

The values of relaxation times are strongly affected by pores geometry and the interfacial property called {\em surface relaxivity}. Indeed, given the boundary conditions, simulations aim to obtain the values of $T_2$ for a given geometry and ralaxivity. Therefore, the input value of $T_2$ in Eq.(\ref{eq1}) will be referred as $T_{2in}$, whereas the output will be labeled simply as $T_2$, or, sometimes, $T_{2out}$. 

 The parameter $D$ is the diffusion coefficient (here taken as a scalar), which, for bulk water, is $D \approx 2\times 10^{-9} m^2/s$ \cite{ref2a}. 

The NMR observable is the time-dependent magnetization, obtained by integrating over the whole volume where the fluid is confined:

\begin{equation}\label{eq1a}
	M(t) = \int_V m(\vec{r},t)dV
\end{equation}In the International System of Units, $M$ is measured in $A/m$ and, therefore, the unit of $m$ is $A/m^4$.

The presence of diffusion and boundaries introduces a geometric  dependence on the solutions of Bloch-Torrey equations, whose importance was correctly picked-up by K.B. Browstein and C.E. Tarr, as explained below. 

Browstein and Tarr established an important relationship between one main NMR observable, namely, the spin-spin relaxation time, $T_2$, and the geometrical parameters of the confining medium. It is important, however, to remark that Browstein-Tarr paper does not solve the Bloch-Torrey equations for a RF pulse sequence, what would be a true NMR experiment. Instead, they assume the free evolution of nuclear magnetization after a $\pi/2$ pulse, under diffusion effects, and propose a multi exponential solution for it, as an {\em ansatz}. 

The truly important insight of Browstein-Tarr was to introduce boundary conditions at the surface of the confining medium which correctly captures the effects of interactions between the fluid nuclear spins with paramagnetic centers in the solid matrix. For that, they introduce a new parameter, $\rho$, called {\em surface relaxivity}, in the so-called Robin boundary conditions:

\begin{equation}\label{eq2}
	\left.D\hat{n}\cdot \nabla m(\vec{r},t) + \rho m(\vec{r},t)\right|_S  = 0
\end{equation}where $\hat{n}$ is the unitary vector normal to the surface boundary at position $\vec{r}$. 

Without the second term, this equation would be a ``zero-flux'' (Neumann) kind boundary condition, meaning that no magnetization ``leaks out''  from the volume $V$. Physically, the main channel of $T_2$-relaxation in a fluid bulk are the fluctuations of dipole-dipole interactions between nuclei. The above boundary condition introduces a new channel of relaxation at the walls of the confining pores, therefore splitting the fluid in two regions: a bulk region and a nanometer-thick layer in contact with the solid surface. 

NMR $T_2$ experiments bring information about {\em porosity}, {\em surface ralaxitivity},  {\em wettability} and {\em permeability},  important petrophysical quantities which impact the potential of fluid extraction from an oil-containing porous rock \cite{ref1}. B. Chencarek et al. published a study of petrophysical quantities measured in synthetic rocks with controlled porosity and permeability \cite{ref5}. 

Unfortunately,  it is not a simple task to experimentally determine the parameter $\rho$, or to predict it theoretically, but it is usually taken in range  10--20 $\mu$m/s, sometimes higher. Recently, M. Nascimento et al. developed a theoretical model which takes explicitly  into account microscopic features of surface relaxivity \cite{ref4}, and E. Lucas-Oliveira and co-workers estimated $\rho$ from $T_2$ distribution and digital rock simulations \cite{ref4a}, reporting values of $\rho$ as high as $50\mu m/s$. One interesting statistical approach for relaxation taking into account paramagnetic impurities in the solid matrix has been reported by J.L. Gonzalez and co-workers \cite{ref4b}.

The fundamental result which links  spin-spin relaxation time, $T_2$, to a porous geometry, and therefore ranks NMR as one of the main petrophysical techniques, can be worked out from the boundary conditions, Eq.(\ref{eq2}), (see Ref. \cite{ref5} for a  deduction). It establishes that, in the so-called {\em fast diffusion regime} (see below), the following expression holds:

\begin{equation}\label{eq3}
	\frac{1}{T_2}=\rho\frac{S}{V}
\end{equation}where $S$ and $V$ are, respectively, the surface and volume of the confining pore. For a spherical pore with radius $a$, this reduces to:

\begin{equation}\label{eq4}
		\frac{1}{T_2}=\rho\frac{3}{a}
\end{equation} This expression is valid as long as the diffusion during an experiment time window $\tau$  is such that \cite{ref3}:

\begin{equation}\label{eq4a}
	\frac{\rho a}{D}\ll 1
\end{equation}This condition establishes the so-called {\em fast diffusion regime}. 

An insightful relation can be obtained by replacing $\rho = a/3T_2$, from Eq. (\ref{eq4}), into Eq. (\ref{eq4a}):
 
 \begin{equation}\label{eq4b}
 \sqrt{DT_2}\gg a 
 \end{equation} The left side is the {\em mean displacement} by diffusion in the time interval $T_2$. This inequality means that, in the fast diffusion regime, the spins will probe the pore walls before the magnetization decays away. This expression yields an useful parameter for numerical calculation.

  From Eq.(\ref{eq3}), we see that, in a natural rock with a distribution of pores sizes, there will be, accordingly,  a distribution of relaxation times. Measuring a $T_2$ distribution, therefore, should allow obtaining the corresponding pore size distribution, characteristic of each rock system. In the presence of strong internal field gradients, however, as it happens to be the case in NMR high-frequency experiments, failing  to reach the fast-diffusion regime adds  ambiguity to the problem, as it will be shown below. 

\section*{Pulsed-NMR COMSOL Programming}
Two previous publications in which COMSOL was used to study NMR systems can be found in Refs. \cite{ref10,ref11}, concerning, respectively, studies of hydrologic parameters of porous rocks, and simulations of pores influence on NMR signal from human cervical vertebra. 

In the present report, COMSOL (license 9200972) is used to solve the Bloch-Torrey and McConnell equations to simulate the following pulsed NMR experiments:
\begin{list}{}{}
	\item -- NMR spectra, obtained from the FFT of the Free Induction Decay, after a $\pi/2$ pulse in spherical domains representing the pores, in different conditions and sizes;
	\item -- Spin-spin relaxation times following the decay of the spin-echo after a Hahn pulse sequence, in both, fast and slow diffusion regimes. Here, an important remark must be done. Hahn sequences are not the best technique to study $T_2$ decay in porous media, particularly in the presence of internal magnetic inhomogeneity. The most used technique is CPMG. However, as stated at the introduction, it is not the purpose here to investigate the effectiveness or advantages of some NMR techniques over others. Magnetic inhomogeneity is neglected in most of the simulations made, and the COMSOL simulation of CPMG is briefly discussed at the concluding remarks;
	\item -- The above two studies in a system of two coupled fluids (McConnell equations).
\end{list}Simulations are made in spherical domains, all the way through. Variations of this, alongside other studies ideas, are suggested at the concluding remarks.

 COMSOL is used to solve Equation (\ref{eq1})  under the boundary conditions of Equation (\ref{eq2}) for the following geometries: first a sphere with radius 20$\mu$m and then a 2$\mu$m one. Effects of slow and fast diffusion regimes on $T_2$ relaxation will be shown, as well as the effects of surface relaxivity. Then, a porous system containing one single 20$\mu$m radius sphere, randomly surrounded by 800 unconnected 2$\mu$m ones is used to simulate an NMR experiment aimed to separate pore sizes by measurement of $T_2$. Finally, the study is repeated for a coupled two-fluid model as a function of the coupling strength, by solving the McConnell equations.

COMSOL Multiphysics is a numerical platform to solve differential equations  using finite elements in arbitrary geometries. The control over geometry is where COMSOL really  shines in the present application, since geometry is so closely related to the petrophysical properties of real oil rocks. 

In the present study the modulus ``Mathematics'' of COMSOL will be used to solve the Bloch-Torrey equations. However, as stated before, it is not the intention of this report to be a manual of COMSOL. There is plenty of introductory material in books, Youtube and web pages about it (see, for instance, https://br.comsol.com/models). The purpose here is solely  to yield a guideline for those with some experience with COMSOL and, hopefully, to motivate the NMR community working with porous media to use this very useful tool.  

After opening COMSOL, follow the clicking path:

\begin{list}{}{}
	\item -- Model Wizard
	\item -- 3D
	\item -- Mathematics
	\item -- PDE interfaces
	\item -- Coefficient form PDE(c)
\end{list} Click ``add''. On the right side window, ``Review Physics Interface'', define ``$m$'' as the field name, ``3'' as the number of dependent variables, and ``$mx,my,mz$'' as the dependent variables labels. On ``Units'' define ``A/m'' for the custom unit dependent variable, and ``A/(m$\cdot$s)'' for the source term. Then click in ``study'' and select ``Time Dependent'' in the ``General Studies'' window.  Then, click in ``Done''. 

COMSOL allows all the physical parameters to be gathered in a table, from which they are read by the program. Table I exemplifies some parameters used in the simulations reported here. 

\newpage
\begin{center}
	Table I  COMSOL parameters used to solve Eq.(\ref{eq1}). Units are mixed, for convenience.
\begin{tabular}{l|c|r}\hline
	Parameter & Value & Description \\ \hline 
	$\Delta$B & $10^{-6}$ T  & Magnetic field offset \\
	$\gamma_n$ & 45 MHz/T  &  Proton gyromagnetic ratio \\
	$\tau$  & $3.4907\mu s$ & Duration of a $\pi/2$ pulse \\
	$B_1$ & 0.01 T & Amplitude of the RF field \\
	$n\omega_1$ & $n\gamma_n B_1$ & Coupling to the RF: $n=1$  ON, $n=0$  OFF \\
	$\Delta\omega$ & $\gamma_n\Delta B$ & Frequency offset \\
	T$_1$ & 500 ms & Spin-lattice input relaxation time \\
	T$_2$ & 50 -- 500 ms & Spin-spin input relaxation time \\
	$D$   & 2 -- 3$\times 10^{-9}$m$^2/s$ & Water bulk diffusion coefficient \\
	$\rho$ & $10^{-6}\mu$m/$s$ & Surface relaxivity \\ 
	$k$ & 0, 10, 100 Hz & Coupling strength for the two-fluid model \\
	$m_0$ & $10^{12}$ A/$m^4$ & Equilibrium magnetization density (arbitrary scaling factor) \\ \hline
\end{tabular}
\end{center}
\newpage
For the $T_2$ simulations using Hahn pulse sequences, the parameter ``$\omega_1$'' must be set ``ON'' during pulses  and ``OFF'' during the intervals. This is done  by simply setting $n$ equal to ``1'' or ``0'', accordingly, in the $\omega_1$ entry line. 

The COMSOL ``Coefficient Form PDE(c)'' modulus presents us with the general equation: 

\begin{equation}\label{eq5a}
	e_a\frac{\partial^2 \vec{m}}{\partial t^2}+d_a\frac{\partial\vec{m}}{\partial t}+\nabla\cdot(-c\nabla\vec{m}-\alpha\vec{m}+\gamma)+\beta\cdot\nabla\vec{m}+a\vec{m}=f
	\end{equation}where $\vec{m}=(m_x, m_y, m_z)^T$ and $\nabla = (\partial/\partial x, \partial/\partial y,\partial/\partial z)$.
	
	To reproduce Eq.(\ref{eq1}) from this general coefficient differential equation, we set $e_a=\alpha=\gamma=\beta=0$, $d_a=(1,1,1)$, $c=(D,D,D)$ and $f=(0,0,m_0/T_1$) ($m_0$ is only a numerical scale factor for graphical convenience). Set the parameter $a$ as the matrix:

	\begin{equation}\label{eq5b}
		a=\left(\begin{array}{ccc}
			1/T_2 & -\Delta\omega & 0 \\
			\Delta\omega & 1/T_2 & -\omega_1 \\
			0 & \omega_1 & 1/T_1
		\end{array}\right)
	\end{equation}

	To introduce the boundary conditions of Eq.(\ref{eq2}) we choose the ``flux/source'' boundary conditions option and set $g=0$ and $q=\rho$ in the equation parameters window. Click in ``Initial Values'' and set $m_z=m_0$ and all the other parameters equal to zero.

After introducing the DE parameters according to Eq.(\ref{eq1}) and Table I, it is necessary to insert three new studies, each of them describing a time interval in the pulse sequence (see below): the first study is a $\pi/2$-pulse. It is followed by a delay $\Delta t$ (the second study). Then, the third study is a $\pi$ pulse, followed by the last study, another $\Delta t$ free-evolution interval. Eq.(\ref{eq1}) is solved in each interval, taking as initial conditions the last values of $m_x, m_y, m_z$  in the previous study. In other to do that, the studies must be  coupled in sequence. 

To add studies, click in the ``Add Studies'' window three times to create the other studies. First and third studies are, respectively, $\pi/2$ and $\pi$ pulses, whereas the second and fourth studies are free-evolution intervals. To couple the first study to the second, expand the second study window and click in ``Step 1'' to open the setting window for this interval. Then, expand the ``Values of Dependent Variables'' window. In ``Settings'' choose ``user controlled''; in ``method'' choose ``solution''; in ``study'' choose ``study 1'' and, finally, in ``time'' choose ``last''. This will instruct the program to take the last point calculated in study 1 (end of $\pi/2$ pulse) as the initial condition for the second study, the free-evolution interval with duration $\Delta t$. Repeat this procedure for the other two intervals. Each study is run in separate. Remember to set $n=0$ in the intervals in which the pulse is OFF (studies 2 and 4) or $n=1$ in which it is ON (studies 1 and 3), before running the program.

After setting all parameters, the next step is to mesh and run the program. As an example, Figure 1(a) shows the output for the magnetization density distribution in a spherical pore with radius $a = 20\mu$m, after the $\pi/2$-pulse, at the instant of time $\Delta t = 2T_2$ (that is, only studies 1 and 2 were run). We can clearly see the different patterns of oscillations of the magnetization density inside the pore. To emphasize the importance of boundary conditions, Figure 1(b) shows the same study, but using the ``zero-flux'' boundary condition (that is, $\rho=0$). In this case the effects of the interactions between the nuclear magnetization and the pore walls, parameterized in $\rho$, are neglected, dramatically modifying the magnetization density distribution inside the pore.

The NMR signal is the time free-evolution of the total magnetization after a pulse sequence, i.e., the integration of $m(\vec{r},t)$ over the porous volume, Eq.(\ref{eq1a}). To obtain that, expand the ``Results'' window and right-click on ``Derived Values''. Then, find the ``Integration'' option and choose ``Volume Integration''. In the setting window click in ``evaluate''. The result will appear as a table. By clicking on the ``table graph'' option, the total magnetization will be plotted as a function of time. Figure 1(c) shows the result for $M_x(t)$ (blue line) and $M_y(t)$ (green line) after a $\pi/2$ pulse applied along $X$. This is the so-called FID signal. The total evolution time was set as $\Delta t=2T_{2in}$. Finally, Figure 1(d) shows the Fourier transform of this signal, which is the NMR spectrum. Unfortunately, the 5.6 version of COMSOL used in the present studies does not have a simple way to calculate FFT. Figure 1(d) was obtained with MatLab (license 1145007) after exporting the table data calculated by COMSOL (there is a click option to export data in the menu).   

\begin{figure}
	\centering
	\includegraphics[width=1.1\linewidth]{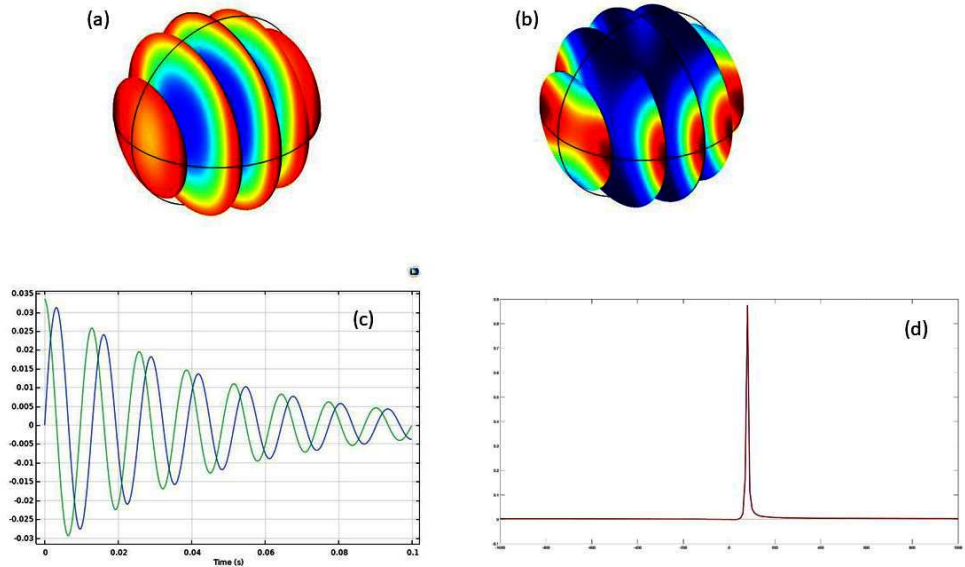}
	\caption{Pore distribution of magnetization density (a and b), FID (c) and NMR spectrum (d), calculated in a spherical pore with radius equal to $20\mu$m: (a) magnetization distribution at the instant of time $\Delta t=2T_2$ following a $\pi/2$-pulse along $X$ with ``flux/source'' boundary condition ($\rho\neq 0$). (b) The same as (a) but with ``zero-flux'' boundary condition ($\rho = 0$). Notice how the boundary conditions change the pattern of oscillations, leaving a layer of surface magnetization in (a). (c) FID (green line $M_y(t)$ and blue line $M_x(t)$). Notice that right after the pulse, $M_x=0$ and $M_y$ is maximum. (d) NMR spectrum (FFT of the FID), with peak at $\Delta\omega/2\pi$.}
	\label{fig:fig1}
\end{figure}

\section*{Rotating Frame and Field Inhomogeneity}
In an experiment, the NMR signal is detected in the so-called ``rotating reference frame''. Mathematically, introducing this frame is rather simple, either classically (Bloch-equations) or quantum mechanically (Schrodinger or Liouville equations) \cite{ref2a}. Electronically, the rotating frame NMR signal detection is accomplished by setting the receiver reference frequency to a value close to the theoretical Larmor frequency $\omega_0=\gamma_nB_0$ of a {\em known NMR probe nucleus}. Depending on the chemical environment of the probing nucleus, there will be a difference between the resonance  and the reference frequencies.  This is called {\em chemical shift} and its effect is to produce oscillations in the FID signal. Besides that, since no magnet is perfectly homogeneous, the local precessing frequency will be a function of the nucleus position inside the magnet. In some modern high-resolution magnets the homogeneity can reach amazing 0.01 ppm. In the present simulations, both effects, chemical shift and magnet inhomogeneity, are mapped into the parameter $\Delta\omega$.

In the rotating frame,  for each ``pack of spins'' which rotates (clockwise) with frequency $+\Delta\omega$ there will be another pack rotating (anticlockwise) with frequency $-\Delta\omega$ (Figure 2). In most situations, one of the contributions to the signal is canceled, remaining only the perpendicular one. 

\begin{figure}
	\centering
	\includegraphics[width=0.7\linewidth]{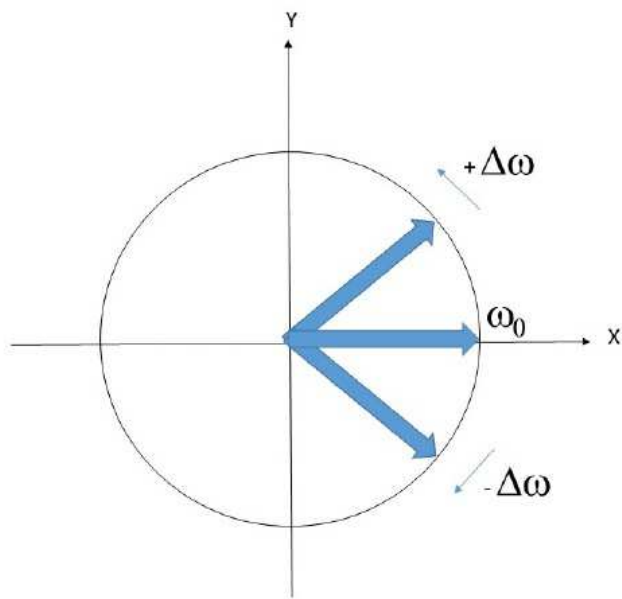}
	\caption{Sketch of the free evolution of nuclear magnetization in the rotating frame, after a $\pi/2$ pulse applied along the axis $Y$. The reference frequency is $\omega_0$. In this frame, neglecting spin-spin interactions, a pack of spins with frequency $\omega_0$ will remain static pointing along $X$. Spin packs which sense a slightly different value of the static field $B_0$, due to field inhomogeneity and chemical shifts will precess symmetrically opposed with frequencies $\pm\Delta\omega$. In most of situations, one the contributions to the NMR signal (in the above scheme, the $Y$ contribution) is canceled.}
	\label{fig:fig2}
\end{figure}

A very  important issue related to NMR technique in porous rocks concerns the magnetic response of the solid matrix to the applied static field. This response increases the local field inhomogeneity, broadening the NMR spectra. Natural rocks will present paramagnetic centers and impurities, and even ferromagnetic  clusters, besides the ubiquitous diamagnetic response. The presence of these elements distort the otherwise nearly-homogeneous field $B_0$ creating strong internal field gradients which affect the dynamics of spins in the liquid phase \cite{ref6}. Each of these magnetic contributions exhibits a characteristic magnetic response which will affect the local field. The overall effect on NMR observables are the NMR line broadening and speeding the relaxation. 

Field inhomogeneity can be considered in COMSOL pulsed NMR simulations. The simplest way is to model the position-dependent magnetic field directly in the Bloch-Torrey equations. Since the information about the local field is in $\Delta\omega$, we can simply model the inhomogeneity using, for instance, a Gaussian distribution (or any other, according to the situation):

\begin{equation}\label{eq7}
	\Delta\omega\longrightarrow\Delta\omega \times e^{-(x^2+y^2)/a^2}
\end{equation}where $a$ is the pore radius. This inhomogeneity is axisymetric about $Z$, the direction $B_0$ is applied. On the symmetry axis, $x,y=0$, the inhomogeneity is, $\Delta\omega$, and at the pore walls it is 1/3 of that, approximately. This model will be used to exemplify the effect of field inhomogeneity on $T_2$ decay in the slow and fast diffusion regimes. 

\section*{Simulation of NMR Spectra in a single pore. Effects of size} The simplest NMR experiment consists on the application of a $\pi/2$ pulse to the magnetization initially at equilibrium along the $Z$ direction. The pulse must be applied perpendicularly to $Z$. The direction $X$ is considered here. The detected signal is the FID (Fig.1(c)) and its Fourier Transform is the NMR spectrum (Fig. 1(d)). These NMR quantities are affected by the geometrical characteristics of the confining medium, as well as the dynamics imposed by the interactions between the fluid and the solid surface.

Figure 3 shows the FIDs and spectra of two isolated pores, one with radius $2\mu$m and the other with radius $0.2\mu$m. For comparison, Figs. 1(c,d) are reproduced ($20\mu$m pore). Apart from the radii, all other parameters were kept the same. We see an obvious change in the decay pattern with the linewidth of spectra changing with the pore radius. The size does not affect significantly the position of the line. For comparison, the calculation show in 3(c) is repeated for $\rho=0$, as shown in 3(d), which becomes similar to 3(a).

\begin{figure}
	\centering
	\includegraphics[width=0.85\linewidth]{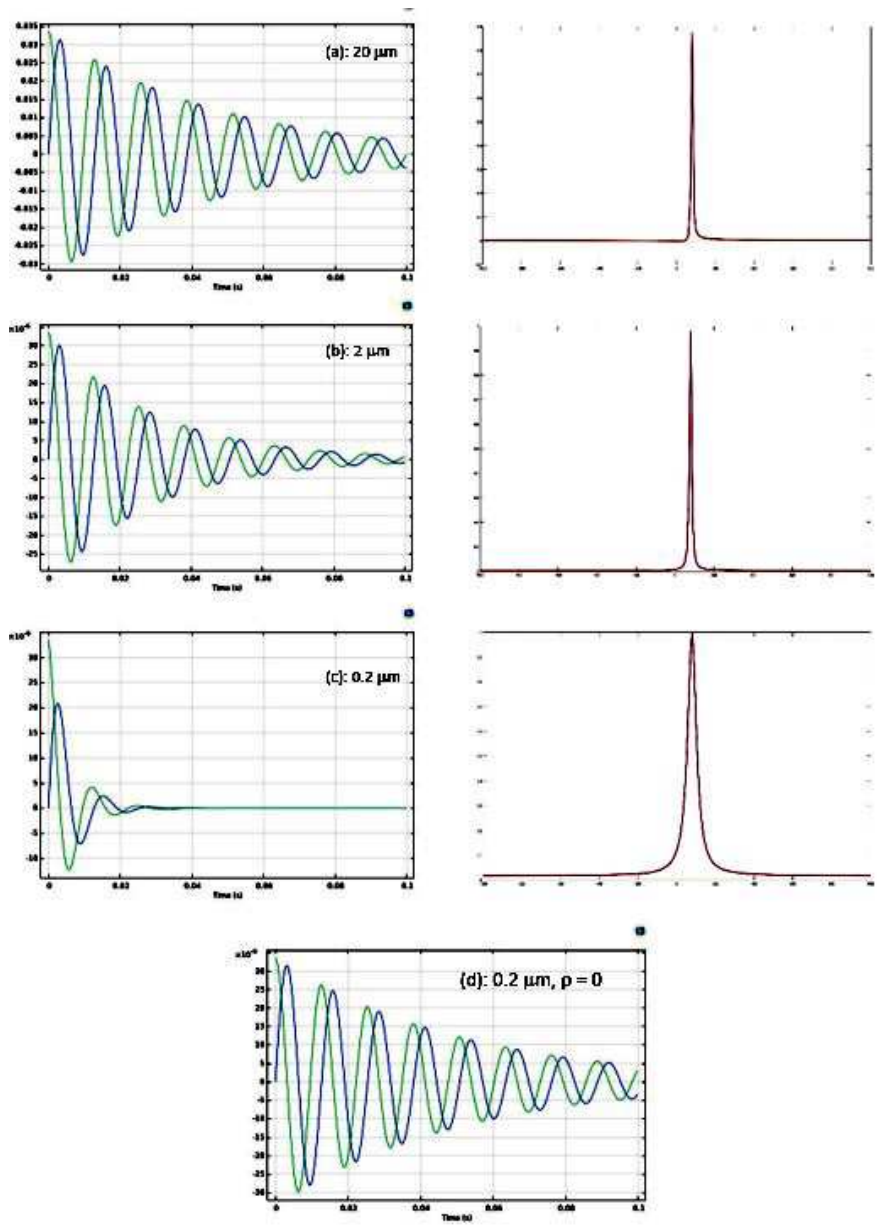}
	\caption{Effects of pore sizes on FID and NMR spectra. These FIDs were calculated using exactly the same parameters of Table I, except for the pores radii, as indicated in the plots. NMR position line does not show a significant change with the pore size, the main effect being the increase of the linewidth. Other shapes, however, may affect position, as well. For a matter of comparison, Fig 3(d) shows the FID calculation, exactly as in (c), but setting surface relaxivity equal to zero.}
	\label{fig:fig3}
\end{figure}

\section*{COMSOL Simulation of spin-spin relaxation in single pores. } The simplest NMR pulse sequence to measure spin-spin relaxation time, $T_2$, is shown in Fig.4. One $\pi/2$-pulse is applied to the equilibrium magnetization, followed by a free-evolution time interval $\Delta t$. Then, a second $\pi$-pulse is applied along the same direction as the first pulse. The echo signal occurs approximately after a further interval $\Delta t$. The amplitude of the Hahn echo is measured as a function of interpulse delay. Each of these intervals is a separated study in COMSOL, but they must be coupled, as explained before. It is important to set correctly the initial and final instants of time for each interval, as well as the step. This is done in the ``Settings'' window of each study. For instance, the second interval $\Delta t$ divided in 500 steps is set as: 

\begin{list}{}{}
	\item -- Start: $\Delta t + 3\tau$ \\
	\item -- Step: $\Delta t/500 $\\
	\item -- Stop: $ 2\Delta t + 3\tau$
\end{list}  The echo amplitude occurs at the instant $2\Delta t + 3\tau $. In the following simulations, $\Delta t$ is varied in fractions of the {\em input} $T_{2in} = 50 ms$, from $T_{2in}/10$ (5 ms) up to $2T_{2in}$ (100 ms), and the echo amplitude is recorded as a function of $\Delta t$. The {\em output} values of $T_2$ follow directly from a exponential fitting, as will be shown below. The fitted $T_2$ carries the effects of pore size, surface relaxivity and diffusion.  It must be emphasized, however, that the NMR sequence most used to measure $T_2$ is CPMG, which allows a fster determination of $T_2$ values and also avoids problems related to diffusion in inhomogeneous internal magnetic fields (see Ref.\cite{ref5}). Hahn sequence is used here because its simplicity to exemplify the NMR simulations in COMSOL. 

Relaxation time $T_2$ is the time scale for the transverse magnetization to loose coherence, due to microscopic spin-spin interactions or other intrinsic effects leading to local field variation inside the material. Another factor which erases the NMR signal is the presence of  magnetic field inhomogeneity. In a porous system, the interplay between $T_2$, inhomogeneity and diffusion is essential to recover pore sizes from $T_2$ measurements. Grossly, we want the spins, through diffusion,  to reach the pore walls before the magnetization decays away. So, for a pore of radius $a$, and a fluid with diffusion coefficient $D$ and relaxation time $T_2$, a  condition to meet this requirement is:

\begin{equation}\label{eq8}
	\sqrt{DT_2} \ge a
\end{equation} For instance, for a sphere with radius $a=2\mu$m, filled with water  ($D = 2\times 10^{-9} m^2/s$) and $T_2=50 ms$, $\sqrt{DT_2} = 10\mu$m and this condition is fulfilled. But if $a=100\mu$m, it is not. Therefore, one can expect that in a real porous rock this condition may be fulfilled  for some pores, but not for others. This brings ambiguity to the interpretation of the results. In Ref.\cite{ref6} this issue is discussed in detail.

\begin{figure}
	\centering
	\includegraphics[width=0.85\linewidth]{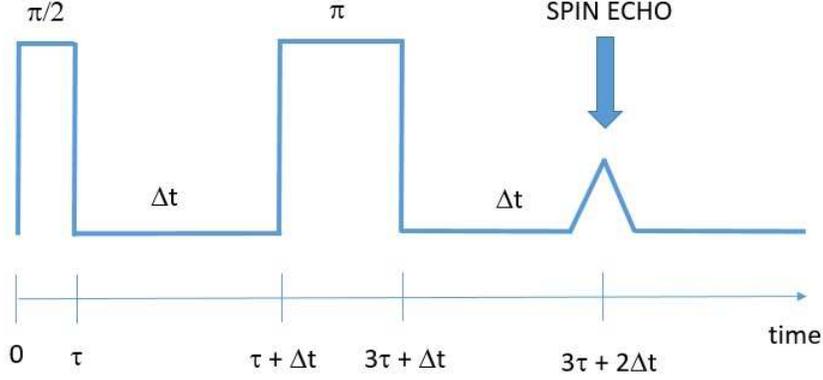}
	\caption{Hahn sequence to measure $T_2$: $\pi/2$ and $\pi$ pulses are applied for various time intervals $\Delta t$ between them. The echo occurs at the instant $3\tau + 2\Delta\tau$ from the first pulse. Echo amplitude decays approximately as $\exp(-2\Delta t/T_2)$. In the present simulations, $\tau = 3.4907\mu$s and $\Delta t$ varies in the interval between $T_2/10$ (5 ms) and $2T_2$ (100 ms)}
	\label{fig:fig4}
\end{figure}
Figure \ref{fig:fig-5-paper} shows the magnetization evolution after the second pulse, for various separation intervals $\Delta t$. The green line is the total component $M_y(t)$ and the blue line $M_x(t)$, the direction the pulses are applied. We see that, at the instant of the echo, $M_x=0$, which means the magnetization is refocused along $M_y$, reaching the maximum at instant $3\tau+2\Delta t$. This is the echo amplitude. 

\begin{figure}
	\centering
	\includegraphics[width=1.0\linewidth]{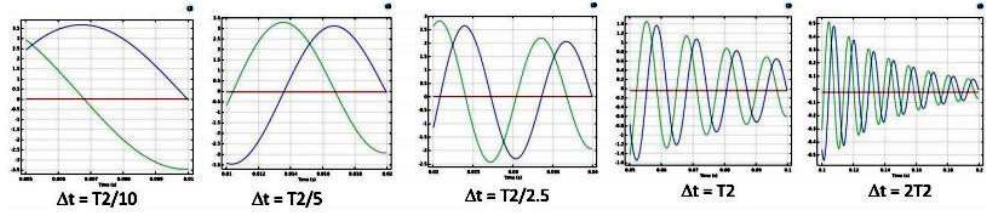}
	\caption{Free evolution of transverse magnetization after a second pulse in a Hahn sequence, for various time intervals between the pulses (see Fig. 4). Green line is $M_y(t)$ and blue line $M_x(t)$. At the instant of the echo, the magnetization is always refocused on the $-y$ direction. This calculation ws done considering a positive offset, $+\Delta\omega$. For the negative counterpart, $-\Delta\omega$ (see Fig. 2), the blue line alone is phase-shifted by 180 degrees.}
	\label{fig:fig5}
\end{figure}

\begin{figure}
	\centering
	\includegraphics[width=0.6\linewidth]{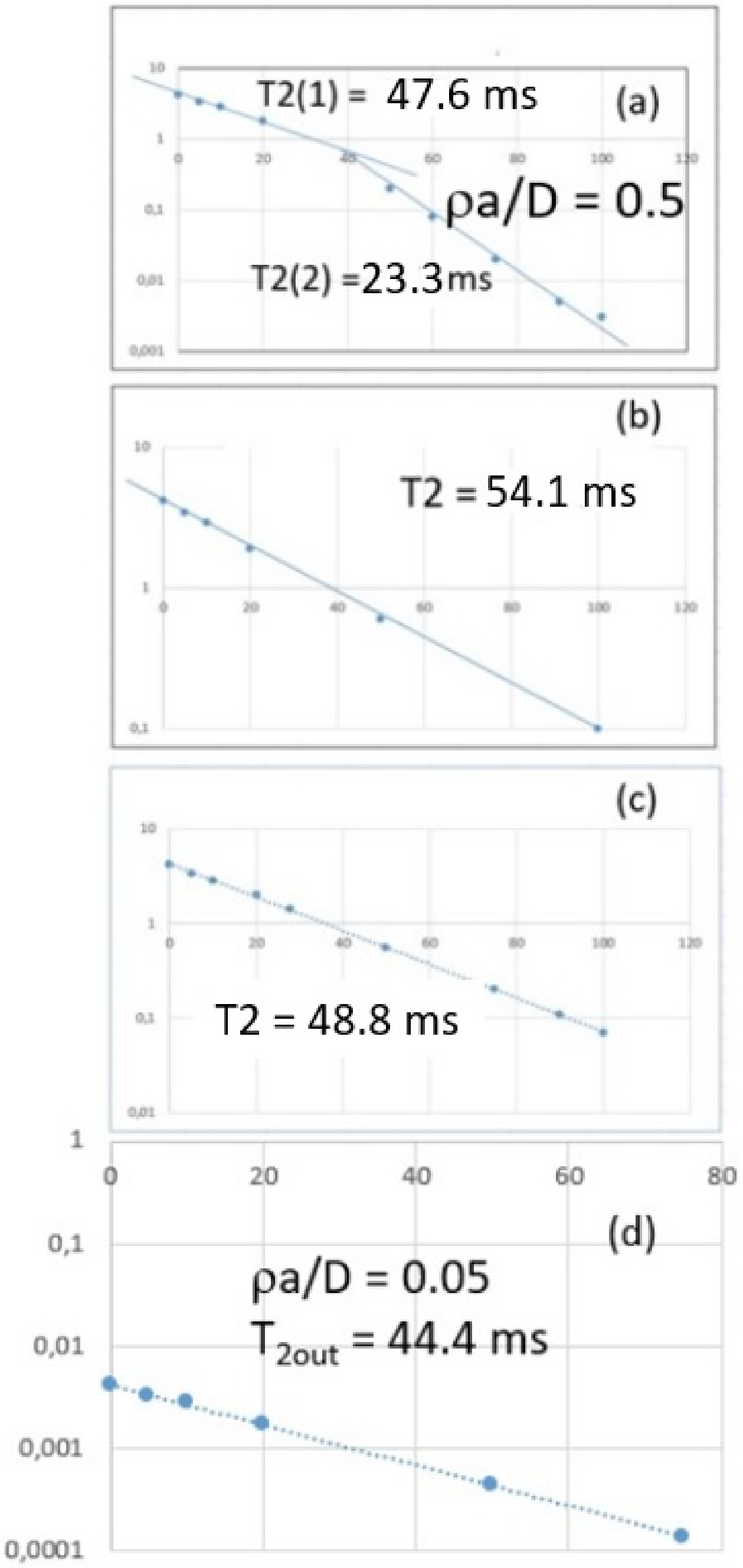}
	\caption{COMSOL simulation of spin-spin decay for a pore with radius $a=100\mu$m. In (a) Gaussian inhomogeneity is included. Slow diffusion regime in the presence of inhomogeneity results in two relaxation regimes. Removing either the inhomogeneity (b) or diffusion (c), results in a single exponential with time decay close to the input value. Surface ralaxivity is $\rho = 10\mu m/s$ in all cases. Figures (a), (b) and (c) were calculated for $\rho a/D = 0.5$. In Figure (d), radius was reduced to $a=10\mu m$, so that, $\rho a/D = 0.05$.}
	\label{fig:fig5}
\end{figure}

Fast diffusion can be reached by, either increasing the diffusion parameter (or $T_{2in}$), or/and decreasing the radius. In order to verify the consistency of the simulations, the relaxation rate for Hahn sequences was calculated as a function of $\rho$ for spheres of three different sizes. Figure 7 shows the results for  $a = 2\mu$m, $a=4\mu$m and $a=20\mu$m. Diffusion was set $D=3\times 10^{-9}m^2/s$ and $T_{2in}$ calculated such as the condition $\sqrt{DT_{2in}}\approx 2a$ is fulfilled in each case. We see that the linearity predicted in Eq.(4) is correctly obtained for the three cases. One important observation is that the larger $\rho$, the more distinguished the pores radii become. At $\rho=10\mu m/s$ the two lines are much closer than at $\rho=50\mu m/s$. Therefore, rocks with large ralaxitivity favor the separation of pore sizes by $T_2$.   

Another consistency test was made by varying the radius and plotting $1/T_{2out}$ vs. $3/a$. In the fast diffusion regime, the angular coefficient is $\rho$. Figure 8 shows the simulation result for an input value of $\rho = 17.5 \mu m/s$. The recovered value, shown in the figure, is very close to that. One experiment like this may just be possible, using modern techniques of lithography to build the porous system, as indicated in Refs \cite{ref7, ref7a} and using NMR microresonators for the detection of very small signals, as exemplified in Refs \cite{ref8,ref8a}.

\begin{figure}
	\centering
	\includegraphics[width=0.7\linewidth]{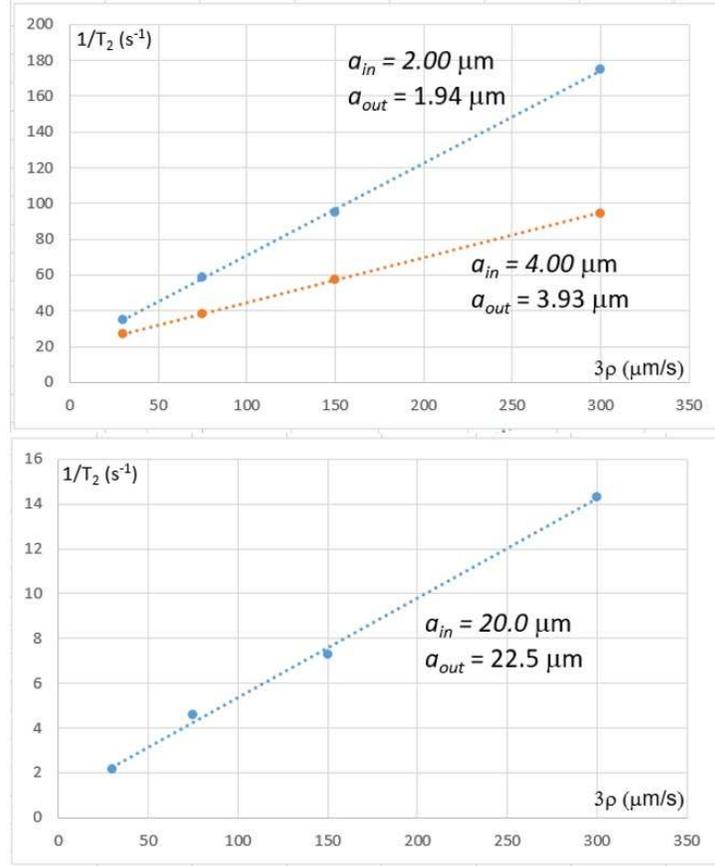}
	\caption{In this simulation, the relaxation rate, $1/T_{2out}$ is obtained for Hahn sequences applied to spherical pores of three sizes, $a = 2\mu$m, $a=4\mu$m and $a=20\mu$m, as a function of relaxivity $\rho$. The idea is to verify the consistency of calculations in the fast diffusion regime, Eq.(\ref{eq3}). The diffusion parameter was kept the same, $D=3\times 10^{-9}m^2/s$, and $T_{2in}$ calculated such as the condition $\sqrt{DT_{2in}}\approx 2a$ was fulfilled in each case. The figure shows the input values of the radii and the recovered ones.}
	\label{fig:fig6}
\end{figure}

In the next Section, these simulations will be applied to a porous medium with two pores sizes. It will be shown that the sizes distribution can be roughly recovered from measurements of $T_2$.

\begin{figure}
	\centering
	\includegraphics[width=0.8\linewidth]{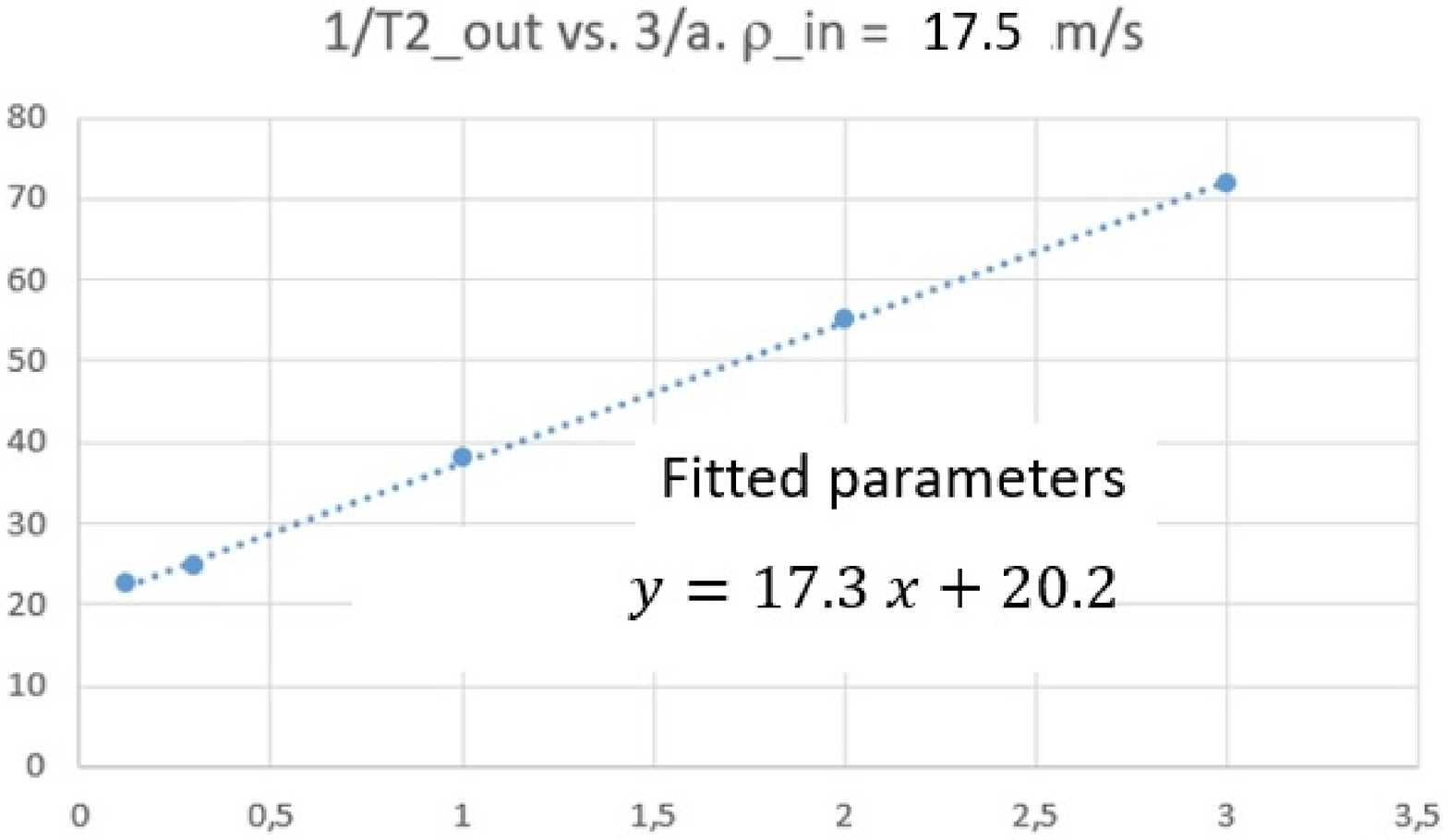}
	\caption{A suggested method to measure the surface relaxivity in artificially built porous media, based on Eq.(\ref{eq3}). For a given surface, spin-spin relaxation time is measured as a function of the radius of the sphere. The horizontal axis is $3/a$. The input value for $\rho$ was $17.5\mu m/s$ and the value obtained from the fit, $17.3\mu m/s$. Artificial porous media can be built using techniques of optical and electron beam lithography \cite{ref7,ref7a} and NMR signals of tiny amounts of sample can be measured using microressonators \cite{ref8,ref8a}.}
	\label{fig:fig7}
\end{figure}

\section*{Simulation of spin-spin relaxation in a ``porous rock'': separating pore sizes by $T_2$ decay} In this Section the simulations described before will be applied to an artificial porous rock containing one single pore with radius $20\mu$m surrounded by 800 randomly distributed others with radius $2\mu$m, each. In order to be able to distinguish the NMR signals of the combined system, the respective volumes of each subsystem must be comparable: $3.4\times 10^{4}\mu$m$^3$ and  $2.7\times 10^{4}\mu$m$^3$, respectively. To build such a system in COMSOL, first a smaller number of random  coordinates were generated in MatLab and inserted in COMSOL. Then, using the symmetry operations of COMSOL, the porous system was constructed, as shown in Fig. 9. In this process, overlap between spheres were allowed, creating more complex geometry of interconnected pores, as shown in the inset of Fig. 9. It also is possible to generate random spheres automatically in COMSOL with different sizes and avoiding overlap, by following an application to simulate...cheese! See Ref.\cite{ref9} for details. It is important to emphasize that the domains of calculation are the interior of the spheres, only. Of course, it is possible to include the solid matrix, and its magnetic response. We will come back to this point in the final remarks. 

\begin{figure}
	\centering
	\includegraphics[width=1.0\linewidth]{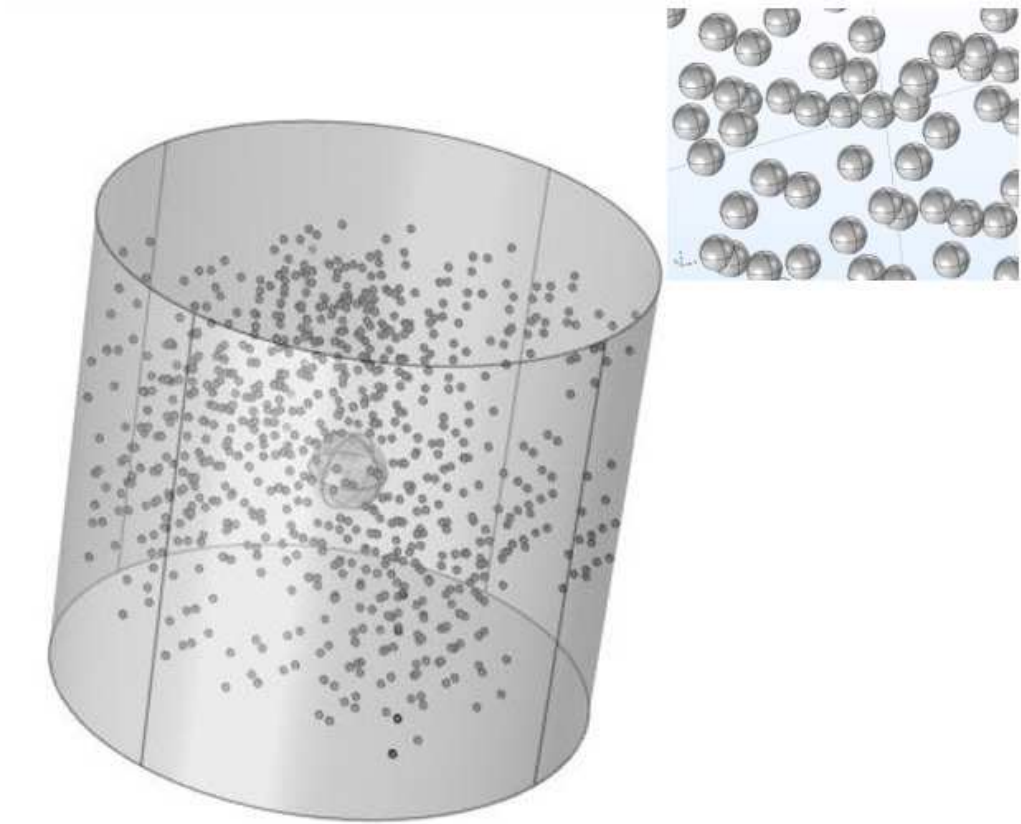}
	\caption{COMSOL construction of a porous rock containing one single spherical pore with radius $20\mu$m, surrounded by 800 identical ones, each with radius $2\mu$m. The respective volumes are $3.4\times 10^{4}\mu$m$^3$ and $2.7\times 10^{4}\mu$m$^3$. The inset shows details of the smaller spheres, which can overlap.}
	\label{fig:fig9}
\end{figure}

The purpose of this Section is to show how a distribution of pore volumes can be obtained from $T_2$ decay experiments. Different pore sizes imply in multiexponential decay in the $T_2$ relaxation. Therefore, one should expect that two pore sizes will produce a two-exponential decay which should be recovered from the measured curve. In some ideal situations the presence of more than one exponential decay can be observed with ``naked eyes''. Figure 10 shows such a case. It is a combination of the decays obtained with COMSOL simulations of a 20 $\mu$m and $1\mu$m spheres. The decays were calculated for individual spheres and the relative weights normalized to 1. This would be the situation of one 20 $\mu$m sphere surrounded by $8000 \times 1\mu$m non-overlapping ones! To build such a system like that of Fig. 9 is not simple, and the calculation becomes too time consuming in an usual workstation. Observe that the recovered values of the radii calculated from the curves are in the correct order of magnitude of the input ones, but with a large error, even for an unusually large value for $\rho$.  

\begin{figure}
	\centering
	\includegraphics[width=1.0\linewidth]{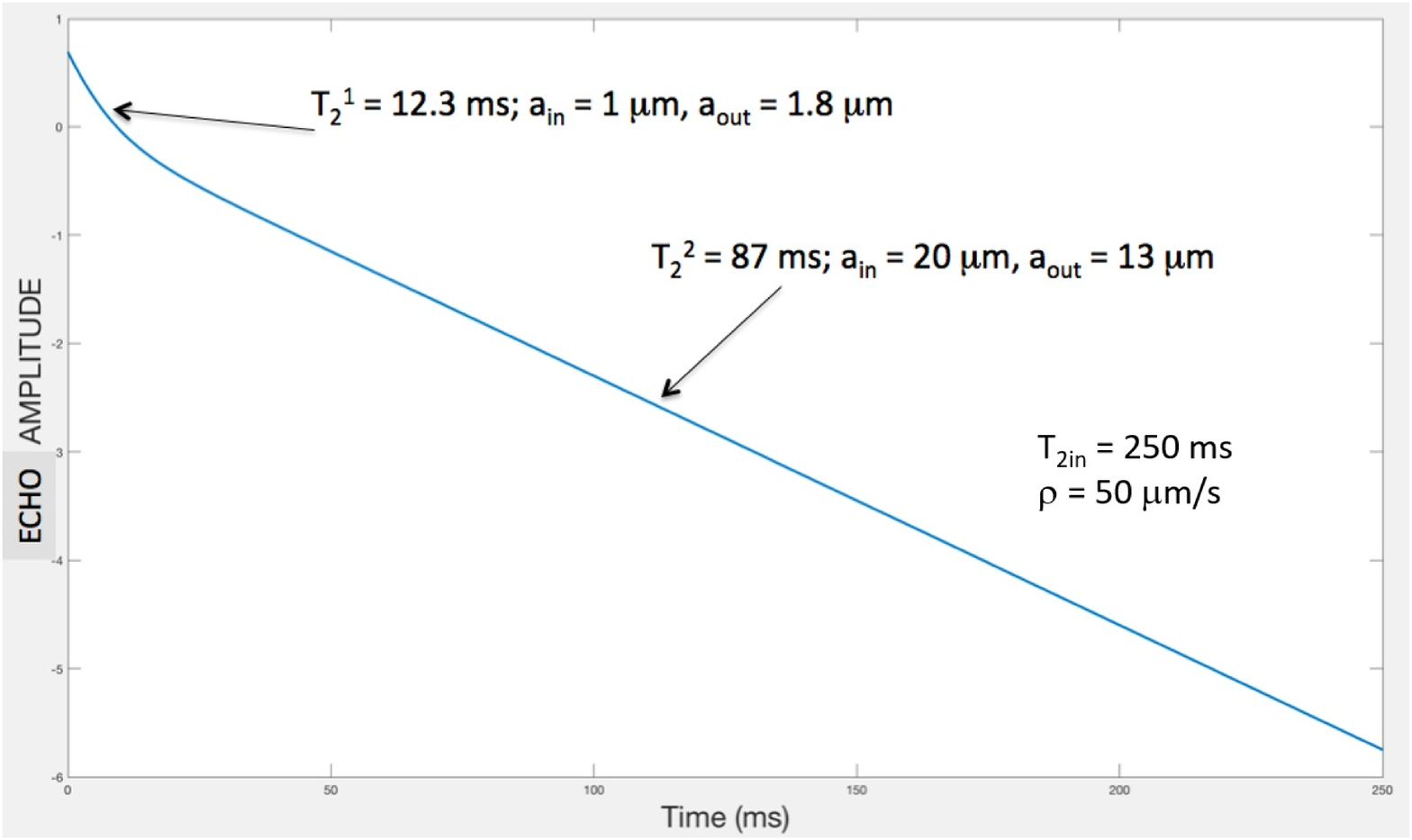}
	\caption{COMSOL simulation of a $T_2$ decay in a system containing one 20 $\mu$m sphere surrounded by $8000 \times 1\mu$m isolated and non-overlapping ones. Two decay constants are clearly visible, but the recovered radii from the measured values show a large error, although they are in the correct order of magnitude. Ralaxivity value was set at $\rho=50\mu m/s$ in order to keep relaxation rates well separated.}
	\label{fig:fig10}
\end{figure}

If we double the size of the smaller spheres and allow overlapping, we have the system depicted in Fig. 9. Here, no longer it is possible to distinguish two decays by eyes.  Fig.11 shows the result of the simulation in linear scale, to emphasize what would be an experimental display. The red line is a two-exponential fit made with MatLab and the black dashed line a one-exponential model. From the figure we can clearly see the two-curves model fits better but, again, although the recovered sizes are in the correct range, the errors are large.  

\begin{figure}
	\centering
	\includegraphics[width=0.7\linewidth]{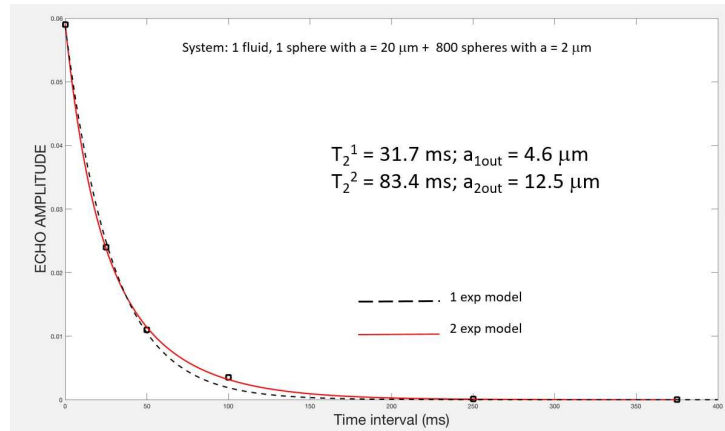}
	\caption{COMSOL simulation of $T_2$ decay in the system of Fig. 9. Squares are the simulated points, the red line a MatLab fit for a two-exponential model and the dashed black line a one-exponential fit. The two-exp model is obviously better, and the recovered sizes are in the correct order of magnitude, but with very large errors.}
	\label{fig:fig11}
\end{figure}

This kind of simulation is close to what would be obtained in laboratory conditions in high-field NMR experiments, in which the level of noise is usually negligible. But in a real rock the situation is much more complicated because there is no control over pore sizes and shapes at all. Different models of fitting are used, as in Ref.\cite{ref5}. In that work it is shown that it is possible to make some sense between sizes and models, at least in a minimally controlled situation. However, if noise gets in, as in the case of current NMR logging technology, then the whole thing may become very doubtful \cite{ref10}! From Fig 11 it is clear that with a very low amount of noise the two curves become blurry and no longer can be distinguished, one from the other. In the next section we extend this study to a two-size pore system containing two fluids. 

\section*{Extension to a two-fluid model}
The ultimate goal of a NMR experiment in porous rocks is to obtain the composition, amounts and mobility of fluids in the porous space, generally composed by a mixture of oil and water. A simple but effective NMR model for coupled fluids with magnetizations $\vec{m}_1$ and $\vec{m}_2$ is given by the so-called McConnell equations \cite{ref11}:

\[
	\frac{\partial\vec{m}_1}{\partial t}-\gamma_1\vec{m}_1\times\vec{B}+\tilde{R}_1(\vec{m}_1-\vec{m}_{10})-k_1(\vec{m}_2-\vec{m}_1)-D_1\nabla^2\vec{m}_1=0
\]
\begin{equation}\label{eq9}
	\frac{\partial\vec{m}_2}{\partial t}-\gamma_1\vec{m}_2\times\vec{B}+\tilde{R}_1(\vec{m}_2-\vec{m}_{20})-k_2(\vec{m}_1-\vec{m}_2)-D_2\nabla^2\vec{m}_2=0
\end{equation} In this model, the two fluids can exchange magnetization through the coupling parameters $k_1$ and $k_2$. Without these terms, the two equations are reduced to a couple of Bloch-Torrey independent equations. In the present COMSOL simulations, we will consider $k_1=k_2=10 Hz$ and $\gamma_1=\gamma_2$, since the nuclear species of interest is, in general, $^1$H in both fluids. 

In order to insert these equations into COMSOL we follow the same procedure as before: first define six dependent variables, the magnetization components of a 6-dimension vector:

\begin{equation}	
	\vec{m}=\left(\begin{array}{c}
	m_{1x} \\
	m_{1y} \\
	m_{1z} \\
	m_{2x} \\
	m_{2y} \\
	m_{2z} \end{array} \right) 	
\end{equation} By opening the terms in Eq.(\ref{eq9}), after some re-arrangement we obtain the matrix form:

\begin{dmath}
	\frac{\partial}{\partial t}
	\left(\begin{array}{c}
		m_{1x} \\
		m_{1y} \\
		m_{1z} \\
		m_{2x} \\
		m_{2y} \\
		m_{2z} \end{array} \right)+\left(\begin{array}{cccccc}
		1/T_{21}+k & -\Delta\omega & 0 & -k & 0 & 0 \\
		\Delta\omega & 1/T_{21}+k & -\omega_1 & 0 & -k & 0 \\
		0 & \omega_1 & 1/T_{11}+k & 0 & 0 & -k \\
		-k & 0 & 0 & 1/T_{22}+k & -5\Delta\omega & 0 \\
		0 & -k & 0 & 5\Delta\omega & 1/T_{22}+k & -\omega_1 \\
		0 & 0 & -k & 0 & \omega_1 & 1/T_{12} + k \end{array}\right)\left(\begin{array}{c}
		m_{1x} \\
		m_{1y} \\
		m_{1z} \\
		m_{2x} \\
		m_{2y} \\
		m_{2z} \end{array} \right)-\left(\begin{array}{cccccc}
		D_1 & 0 & 0 & 0 & 0 & 0 \\
		0 & D_1 & 0 & 0 & 0 & 0 \\
		0 & 0 & D_1 & 0 & 0 & 0 \\
		0 & 0 & 0 & D_2 & 0 & 0 \\
		0 & 0 & 0 & 0 & D_2 & 0 \\
		0 & 0 & 0 & 0 & 0 & D_2 \end{array}\right) + \nabla^2\left(\begin{array}{c}
		m_{1x} \\
		m_{1y} \\
		m_{1z} \\
		m_{2x} \\
		m_{2y} \\
		m_{2z} \end{array} \right) =\left(\begin{array}{c}
		0 \\
		0 \\
		m_{01}/T_{11} \\
		0 \\
		0 \\
		m_{02}/T_{12} \end{array} \right)
\end{dmath} where $T_{2(1,2)} = 250 ms (50 ms)$ is the input value for the spin-spin relaxation time of fluid 1(2). The same notation applies for the spin-lattice $T_{1(1,2)} = 500 ms$. $D_1 = 3\times 10^{-9}m^2/s$ and $D_2=D_1/2$ are the respective diffusion coefficients and $m_{01}=m_{02}=10^{12} A/m^4$ the equilibrium magnetization. In order to simulate different chemical shifts and distinguish the NMR spectral lines, we set the frequency offsets as $\Delta\omega$ and $5\Delta\omega$ for fluids 1 and 2, respectively.  

Let us first investigate the NMR spectra generated by the model. Fig. 12 shows the FIDs and respective FFTs, calculated using the parameters of Table I, for a spherical pore with radius $20\mu$m, for $k=10$ Hz, $k=100$Hz and $k=1$kHz. We see that, as the coupling strength increases, the initially separated lines become a single one. As far as the NMR spectrum is concerned, in this limit, the mixture behaves as a single fluid, and the two components cannot be separated by looking at the spectrum. The same happens to relaxation. 

\begin{figure}
	\centering
	\includegraphics[width=1.1\linewidth]{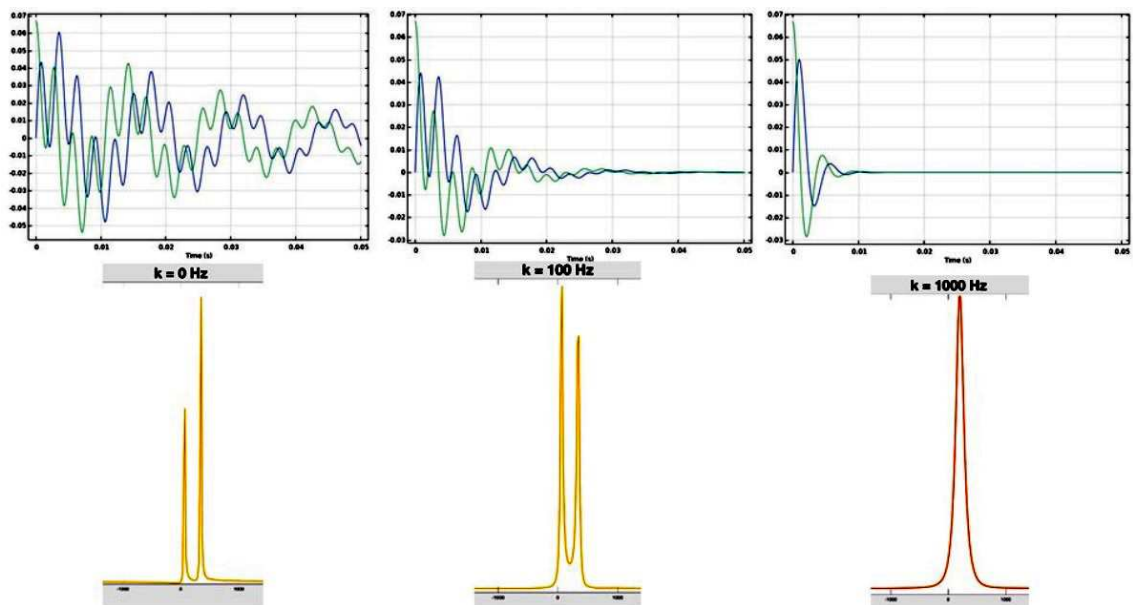}
	\caption{Solutions of McConnell equations for two fluids with different coupling parameters for a single spherical pore with $a=20\mu m$. On top, the FIDs and on the bottom, the calculated spectra. The general effect of the coupling is to merge the two lines into a single broader one, and the relative amount of fluids can no longer be obtained from the spectral areas.}
	\label{fig:fig12}
\end{figure}
\pagebreak
\begin{figure}[h]
	\centering
	\includegraphics[width=0.9\linewidth]{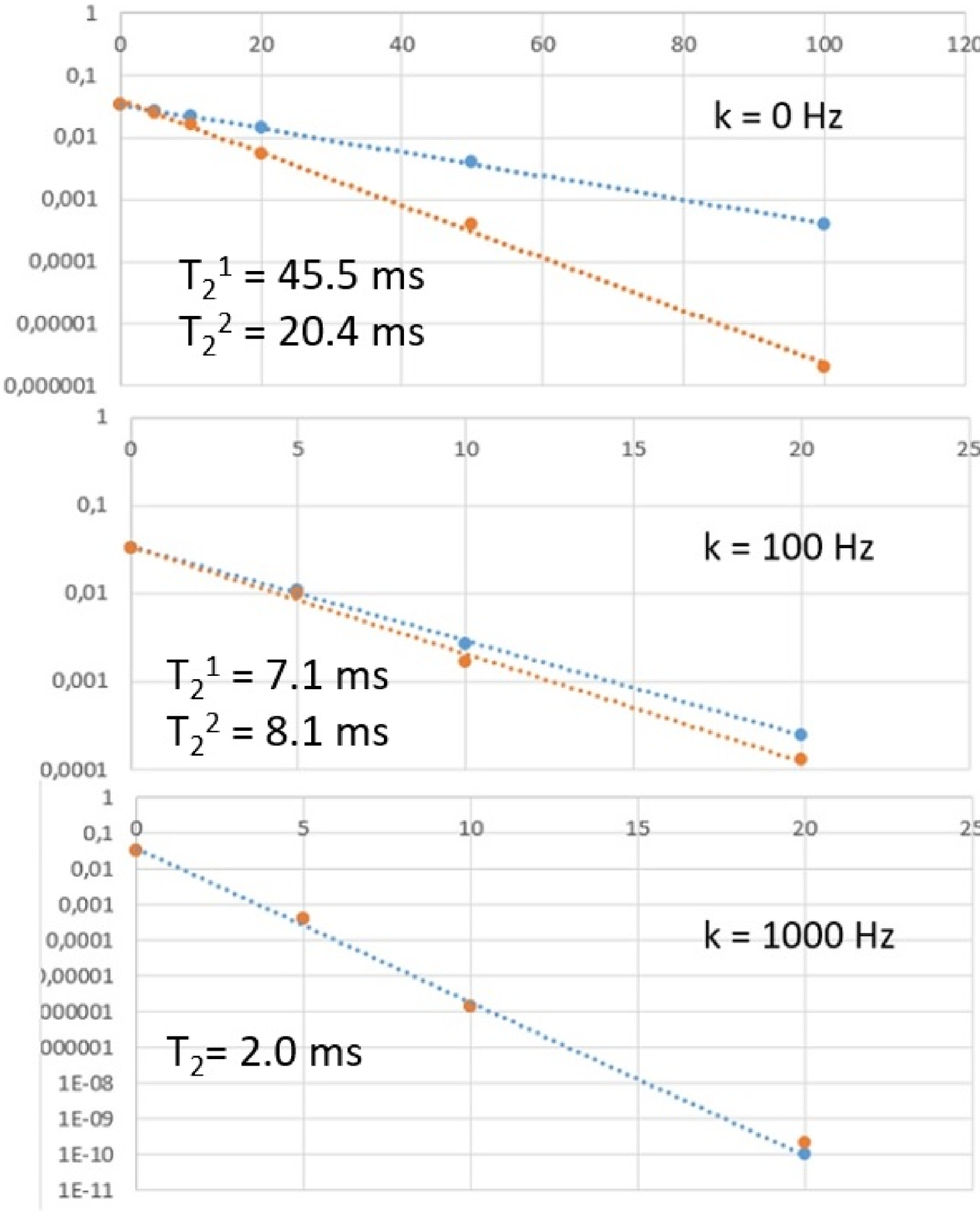}
	\caption{Spin-spin decays for two fluids, as a function of the coupling strength, calculated in a spherical pore with $a=20\mu$m. The effect of coupling is to merge the two lines into a single one, making them indistinguishable The calculation parameters are given in Table I.}
	\label{fig:fig13}
\end{figure}

Figure 13 shows the calculated $T_2$ decay for $k = 0, 100, 1000$Hz, and the respective values of fitted $T_2$ parameters. Notice that for $k = 0$, the time between pulses was extended with respect to the others in order to emphasize the decays of the uncoupled fluids. One observes that, as the coupling increases, the two decaying lines merge into a single one.  

Let us move now for the final application considered in this report: the calculation of $T_2$ decay for the two-fluid model in the geometry shown in Fig. 9.

Figure 14 shows the results of calculation of $T_2$ decay for two coupled fluids in the system of Fig. 9. For this it was considered $k=10 Hz$, and, therefore, the two lines are resolved allowing to follow the individual decays. As before, the red line is a two-exponential model and the black dashed line a single-exponential fit. We can clearly see that two lines can be distinguished in the slower decay (i.e., for fluid 1), but they collapse in a single decay in the faster one (i.e., for fluid 2). The values of decay constants and respective output radii are shown in the figures. Again, the correct order of magnitudes for the pores sizes are obtained from the two lines.  

\begin{figure}[h!]
	\centering
	\includegraphics[width=1.0\linewidth]{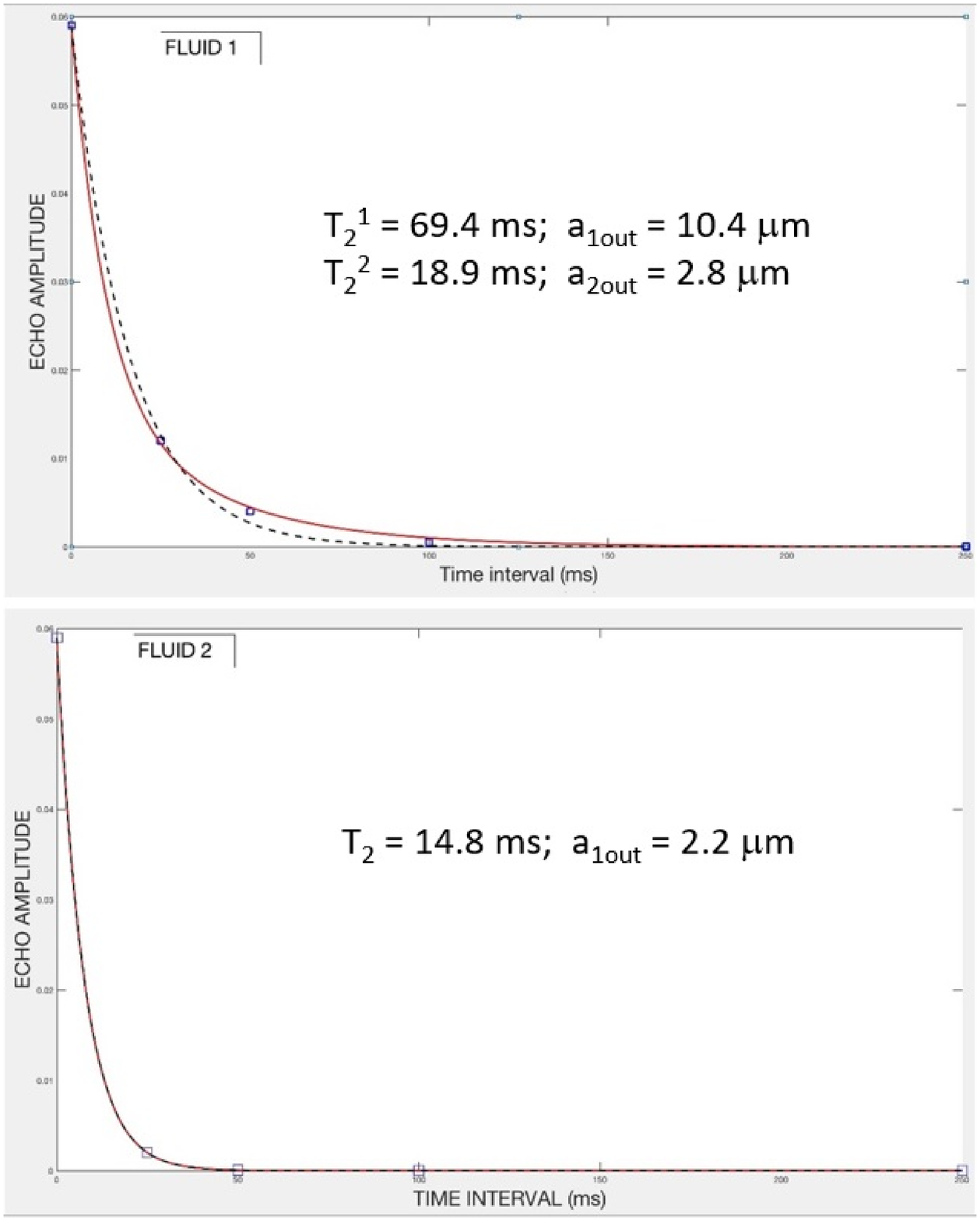}
	\caption{Spin-spin decays in the two fluids model, for the system of Fig. 9. On top: red line is a two-exp fit and black dashed line a one-exp fit. Relaxation times and derived pores radii correspond to the red line. On the bottom: relaxation decay for fluid 2; one and two-exponential models collapse on the same curve and correspond with smaller pores contribution. Results are comparable to those of one fluid case shown in Fig. 11.}
	\label{fig:fig14}
\end{figure}

\section*{Final remarks and conclusions} COMSOL has proved to be a very useful tool to simulate NMR phenomena in porous media with simple geometry. Measurement of relaxation times  with Hahn sequences is only a very basic technique among a plethora of NMR experimental resources. The results  reported here show that, even under absolute control of geometry and noiseless condition, $T_2$ obtained from Hahn sequences can yield, at the best, a rough idea of pore size, and opens the way for further simulations:

\begin{list}{}{} 
	\item \textbf{Variation of geometry} - Geometry is a key element in petrophysical properties of natural rocks. The possibility to control the pore geometry in COMSOL is  truly an advantage of the method. The simulations presented in this report show that correct order of magnitudes of pore sizes can be obtained from $T_2$ decay experiments. One natural way would be to add spheres with other sizes to the problem.  In the limit, it may be possible to load a real porous geometry, or even an entire porous rock image,  obtained from X-ray microtomography, or other tomographic techniques. However, computational capabilities become an important issue if geometry gets too complex. The simulations of $T_2$ decay in the two-fluid model section took 80 minutes, on average, for each point, in a HP Z840 workstation, even without  mesh refinement. Complex geometries may require finer or extra-fine mesh, and time of calculation is bound to grow considerably;
	\item \textbf{Adding throats} - In real porous media, pores are not isolated, but connected by throats. The addition of throats connecting  pores is straightforward in COMSOL. A study of diffusion of spins under field gradients between pores, through throats can reveal interesting aspects of fast and slow diffusion regimes; 
	\item \textbf{Different pulse sequences} - Only Hahn sequences were exploited in this report, for simplicity. But the most used sequence to measure $T_2$ in porous media is CPMG, which can also be simulated in COMSOL. For instance, apply a $\pi/2$ pulse along $X$ $(\phi = 0)$ in Eq. (1), and a sequence of $\pi$ pulses along $Y$ $(\phi = \pi/2)$ with short intervals between them.  One of the most powerful aspects of NMR is the possibility to engineer specific pulse sequences (and pulse shapes!) which corrects dephasing caused by field inhomogeneity of all kinds;
	\item \textbf{Magnetic field inhomogeneity} - Internal magnetic field distributions strongly affect the NMR observables. Different models of magnetic inhomogeneities can be simulated in COMSOL and their effects on lineshapes and relaxation studied, keeping the control over all other parameters. Specific pulse sequences can be built to test corrections of the NMR signal for different models. Here, one important possibility is to \textit{couple physics} with COMSOL. The exact magnetic response of a material can be calculated in one section, and the results can be loaded in a $T_2$ simulation section, allowing to introduce realist magnetic inhomogeneities in the simulations, as shown in Fig. 15;
	
\begin{figure}
	\centering
	\includegraphics[width=1.1\linewidth]{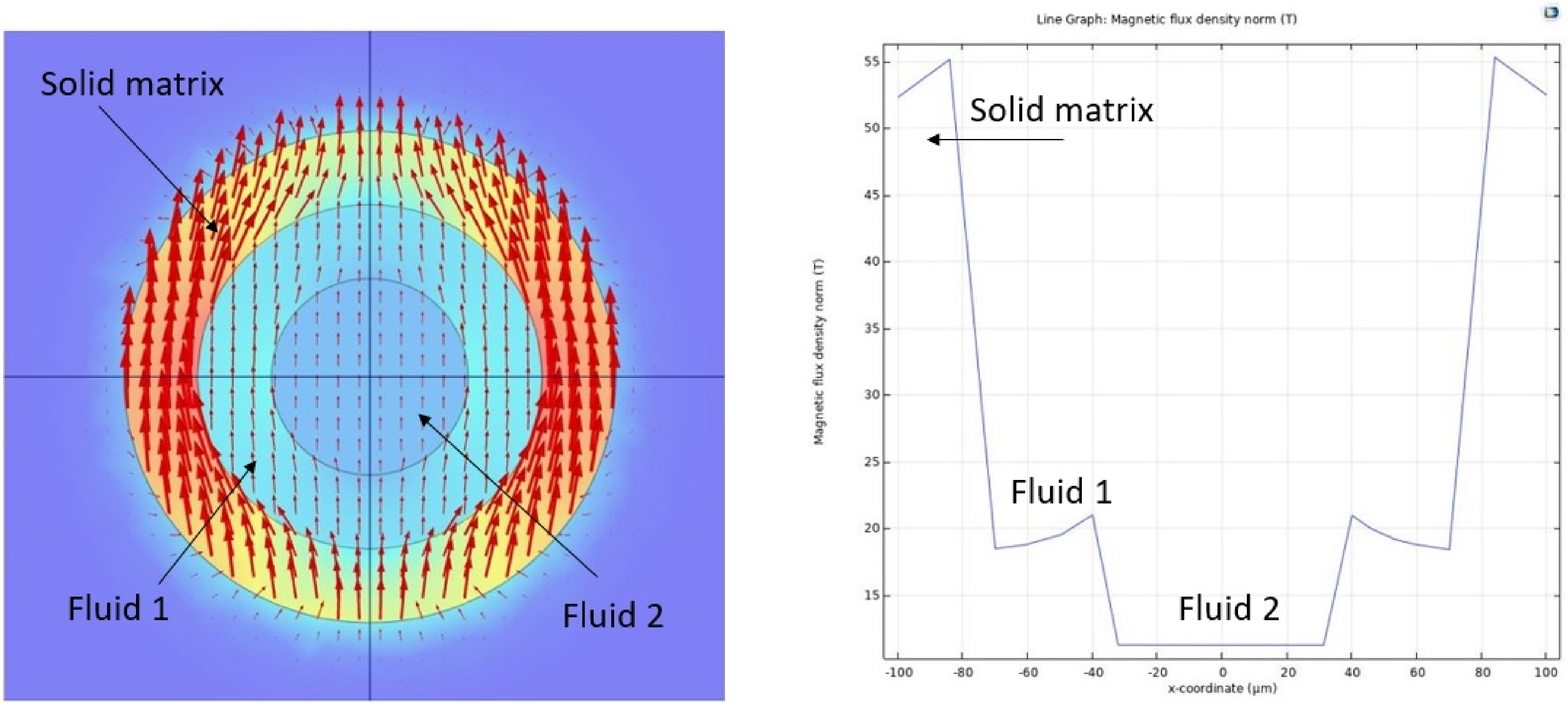}
	\caption{Cross section showing the magnetic induction distribution inside a pore with radius 100 $\mu m$, containing two fluids. The field is applied along the vertical axis. On the right side the field profile along the horizontal axis. COMSOL allows to couple this study as input for solving the Bloch-Torrey equations.}
	\label{fig:fig15}
\end{figure}

	\item \textbf{Time-dependent diffusion} - NMR techniques based on  time-dependent diffusion are powerful alternatives to simple $T_2$ measurements techniques. Introducing a time dependent diffusion in the Bloch-Torrey-McConnell equations and solving them in COMSOL for different geometries and conditions, can possibly lead to important insights, as demonstrated in Ref. \cite{ref12}; 
	\item \textbf{Echo shapes} - The approach used here considered only one isochromat, that is, a single pack of spins precessing with the same offset frequency  $\Delta\omega$. A real echo signal is formed by the addition of various contributions with different precessing frequencies. The final echo shape brings information about field inhomogeneity and efficiency of pulse sequences to correct dephasing. Studying echo shapes in COMSOL simulations as a function of magnetic field inhomogeneity, porous geometry and pulse sequences can be very insightful;  
	\item \textbf{Fluid conformation} - The two-fluid model simulated here did not take into account an important aspect of fluid mixtures in porous rocks: conformation. One possible interesting study with COMSOL is to consider layers of oil and water in contact with a solid surface in different conformations and geometries. 
	\item \textbf{Adding noise} - Noise is intrinsic to any measured data. One of the advantages of high-field NMR in the study of porous media is the extreme sensitivity and low-noise condition. On the other hand, in logging conditions, NMR signals are very noisy, what makes the analysis of signal much more difficult. Noise can be introduced in COMSOL simulations using MatLab routines. This may allow simulating NMR experiments in logging conditions.      
\end{list} The ultimate simulation of NMR phenomena in porous rocks using COMSOL would be to solve the Bloch-Torrey-McConnell equations in a realistic porous digital rock, under different pulse sequences, parameters and techniques, taking into account all the above aspects (and many others!). That would require a considerable computational capability, but would return a very powerful tool   to exploit this important technique for the oil industry and other sectors of science and technology. 
\newpage

\textbf{Acknowledgments}: The author wish to thank Prof. J.P. Sinnecker from CBPF and Dr. A.L. Roman from COMSOL LTd, for their assistance with COMSOL programming.  My acknowledgment to Dr. B. Chencarek, Dr. M. Nascimento, Dr. L. Cirto, Dr. A.M. Souza, Dr. M.S. Reis, Dr. V.L.B. de Jesus, Prof. J.L.G. Alfonso, and Prof. J.C.C. Freitas, for their comments, criticisms and suggestions to this manuscript and the Brazilian Oil Company, Petrobras, for the financial support (Projects 2019/00062-2,  2017/00486-1 and 2019/00062-2), CNPq and Faperj (CNE E-09/2019).

\end{document}